\documentclass{rps-apssra12}

\usepackage{bstnotations}
\renewcommand{\Esp}[1]{ \textrm{I\!E}\left[ #1 \right]}
\renewcommand{\Espe}[2]{{\textrm{I\!E}}_{#1}\left[#2\right]}

\renewcommand{\Nn}{\textrm{I\!N}}
\renewcommand{\Pp}{\textrm{I\!P}}
\renewcommand{\Rr}{\textrm{I\!R}}
\renewcommand{\Prob}[1]{{\textrm{I\!P}}\left( #1 \right)}

\usepackage{natbib}

\def\papername{\jobname}

\copyline{Fifth Asian-Pacific Symposium on Structural Reliability and
  its Applications (5APSSRA).\\
 {\itshape Edited by} Phoon, K. K., Beer, M., Quek, S. T. \& Pang, S. D.}{2011}{981-973-0000-00-0}

\begin{document}

\title{Meta-models for structural reliability and uncertainty quantification}

\author{BRUNO SUDRET}

\address{Universit\'e Paris-Est, Laboratoire Navier (ENPC/IFSTTAR/CNRS),
  \'Ecole des Ponts ParisTech, France \email{bruno.sudret@enpc.fr}}

%\address{$^{2}$Group, University/Organization,
%Country. \email{secondauthor\_id@domain\_name.com}

\begin{abstract}
  A meta-model (or a surrogate model) is the modern name for what was
  traditionally called a response surface. It is intended to mimic the
  behaviour of a computational model $\cm$ (\eg a finite element model
  in mechanics) while being inexpensive to evaluate, in contrast to the
  original model $\cm $ which may take hours or even days of computer
  processing time. In this paper various types of meta-models that have
  been used in the last decade in the context of structural reliability
  are reviewed.  More specifically classical polynomial response
  surfaces, polynomial chaos expansions and kriging are addressed. It is
  shown how the need for error estimates and adaptivity in their
  construction has brought this type of approaches to a high level of
  efficiency. A new technique that solves the problem of the potential
  biasedness in the estimation of a probability of failure through the
  use of meta-models is finally presented.
\end{abstract}


%%%%%%%%%%%%%%%%%%%%%%%%%%%%%%%%%%%%%%%%%%%%%%%%%%%%%%%%%%%%%%%%%%%%%%%%%%%%%%%%%%%%%%%%%%%%%%%%%
\section{Introduction}

The coupling of mechanical models and probabilistic approaches has gain
a lot of interest in the literature in the last 30~years. More
specifically the field of structural reliability has emerged in the mid
70's in order to provide methods for assessing structures by accounting
for uncertainties in both the models and the parameters describing the
structures (\eg geometrical parameters, material properties and applied
forces). Structural reliability is nowadays a mature field with
industrial applications in domains ranging from civil, environmental,
mechanical \& aero-space engineering \citep{Ditlevsen1996}. It has given
a sound basis for the semi-probabilistic structural codes such as the
Eurocodes and has lead to {\em performance-based} engineering.

In the last 10~years the field of stochastic spectral methods has blown
up based on the pioneering work by \citet{Ghanembook1991}. It has become
a field in itself which is at the interface of computational physics,
statistics and applied mathematics as shown in the literature on this
topic, which is published in scientific journals of these various
domains, namely {\em Prob. Eng.  Mech., Structural Safety, Reliab. Eng.
  Sys. Safety, J. Comput. Phys., Comput.  Methods Appl. Mech. Engrg.,
  Comm. Comput. Physics} on the one hand, {\em SIAM J. Sci. Comput.,
  SIAM J.  Num. Anal., J. Royal Stat.  Soc.}  on the other hand, among
others.

The large majority of computational methods associated with structural
safety and uncertainty quantification rely upon {\em repeated calls} to
the underlying computational model of the structure or system. For
instance Monte Carlo simulation is based on the sampling of the input
parameters according to their distribution, and the evaluation of the
model response (or system performance in the context of reliability) for
each realization. The large number of calls that is required for
precise predictions is usually not compatible with costly computational
models such as finite element models, even when high-performance
computing platforms are at hand. This has opened a new field of research
broadly called {\em meta-modelling}. This is the goal of this paper to
provide some overview of meta-modelling techniques with a focus on their
use in structural reliability.

The paper is organized as follows. Section~2 describes the ingredients
of uncertainty quantification and recalls the limitations of Monte Carlo
simulation for structural reliability analysis. Section~3 describes the
general philosophy of meta-modelling and addresses in details three
types of meta-models: polynomial response surfaces, polynomial chaos
expansions and kriging. Section~4 shows how a meta-model can be used in
the context of importance sampling in order to provide unbiased
estimators of a probability of failure. Finally Section~5 gathers some
application examples.

%%%%%%%%%%%%%%%%%%%%%%%%%%%%%%%%%%%%%%%%%%%%%%%%%%%%%%%%%%%%%%%%%%%%%%%%%%%%%%%%%%%%%%%%%%%%%%%%%
\section{Problem statement}
\label{sec:2}
\subsection{Computational model}
\label{sec:21}
Let us consider a mechanical system whose behavior is modelled by a
set of governing equations, \eg partial differential equations
describing its evolution in time. After proper discretization using \eg a
finite element (resp. finite difference) scheme, and using some suitable
solving scheme, the computational model may be cast as follows:
\begin{equation}
  \label{eq:001}
  \ve{y} = \cm(\ve{x})
\end{equation}
In this equation $\ve{x} \in \cd_{\Ve{X}} \in \Rr^{M}$ is a vector
describing the {\em input parameters} of the model. It usually gathers
the parameters describing the geometry of the system, the constitutive
laws of the materials and the applied loading. Vector $\ve{y} \in
\Rr^{Q}$ gathers the {\em response quantities} which may contain:
\begin{itemize}
\item the displacement vector or selected components of the latter;
\item the components of the strain (resp. stress) tensor at specific points;
\item internal variables (\eg plastic strain, damage variables, etc.);
\item a combination of the latter, at a specific point-in-time or at
  various time instants.
\end{itemize}

In the sequel the {\em computational model} $\cm$ is considered as a
black box, which is only known point-by-point: if a given set of input
parameters $\ve{x}_0$ is selected, running the model provides a unique
response vector. It is also assumed that this model is purely
deterministic: running twice the model using the same input vector will
yield exactly the same output.

Note that in practice evaluating such a computational model may be almost
instantaneous if some analytical solution to the constitutive equations
exists, whereas it can take hours on high performance computers if it
results from a large size finite element model (or a workflow of chained
models).

In the sequel the various methods reviewed for uncertainty
quantification and reliability analysis consider that the model cannot
be modified by the analyst but only run for a set of input vectors.
These methods are termed {\em non intrusive} in the context of
uncertainty propagation.

\subsection{Probabilistic model }
\label{sec:22}
Let us consider that the uncertainties in the model input
parameters are modelled by a random vector $\ve{X}$ with support
$\cd_{\Ve{X} } \in \Rr^M$ and prescribed probability density function
$f_{\ve{X}}(\ve{x})$. It is beyond the scope of this paper to describe
in details how such a probabilistic model can be built from the
available information (resp. data). However the following guidelines may be
found in the literature:
\begin{itemize}
\item When some natural variability of a parameter is evidenced through
  a set of measured values, the classical approach consists in using
  {\em statistical inference methods} in order to fit the best
  distribution selected among one or several families (\eg Gaussian,
  lognormal, Gamma, Weibull, etc.) using \eg the maximum likelihood
  principle. The goodness-of-fit shall be checked using appropriate
  tests \citep{Kendall1999}. Eventually {\em the} best distribution may
  be selected using criteria such as the Akaike (resp. Bayesian)
information criterion \citep{Akaike73, Schwartz1978}.
\item In contrast when no data is available, {\em expert judgment}
  should be resorted to. The available information may be
  ``objectively'' modelled using the {\em principle of maximum entropy}
  \citep{Jaynes1982, Kapur92}.  Guidelines such as the {\em JCSS
    probabilistic model code} \citep{Vrouwenvelder97} are available in
  the literature for modelling loads and material properties in civil
  engineering (see updated versions of the code at
  \texttt{http://www.jcss.byg.dtu.dk}).
\item In situations where few data is available, a prior expert judgment
  can be combined with measurements using the framework of Bayesian
  statistics.
\end{itemize}

\subsection{Structural reliability}
\label{sec:2.3}
In structural reliability analysis the performance of the system is
mathematically described by a {\em failure criterion}
which depends on:
\begin{itemize}
\item the (uncertain) mechanical response of the system, say
  $\cm(\ve{X})$;
\item possibly additional deterministic parameters $\ve{d}$ (\eg a
  codified threshold) or random variables $\ve{X}'$ (\eg uncertain
  resistance)
\end{itemize}
The failure criterion is mathematically represented by a {\em limit
  state function} (also called {\em performance} function) which is
conventionally defined as follows:
\begin{itemize}
\item The set of parameters $\{ \Ve{x}, {\Ve{x}'}, \ve{d}\}$ such that
$g \prt{\cm({\Ve{x}}), \Ve{x}', \ve{d} } > 0$ defines the {\em safe
domain} $\cd_s$.
\item The set of parameters $\{ \Ve{x}, {\Ve{x}'}, \ve{d}\}$ such that
$g \prt{\cm({\Ve{x}}), \Ve{x}', \ve{d} }\le 0$ defines the {\em failure
  domain} $\cd_f$.
\item The {\em limit state surface} corresponds to the zero-level of the
  $g$-function.
\end{itemize}

Gathering all the parameters into a single notation $\ve{x} \in
\cd_{\Ve{X} } \subset \Rr^M$ for the
sake of simplicity, the probability of failure of the system is defined
by:
\begin{equation}
  \label{eq:002}
  P_f= \textrm{I\!P} \bra{g(\Ve{X} ) \le 0} = \int_{\cd_f = \acc{\ve{x} : g(\ve{x}) \le 0}} f_{\ve{X}}(\ve{x}) \, d \ve{x} 
\end{equation}
The main difficulty in evaluating the probability of failure
in Eq.(\ref{eq:002}) leads in the fact that the integration domain is
defined {\em implicitely}. Moreover the dimension of the integral is equal to
the number of uncertain parameters $M$ which is usually large. 

\subsection{Classical computational methods}
\label{sec:2.4}
\subsubsection{Monte Carlo simulation}
\label{sec:2.4.1}
Monte Carlo simulation is the basic and universal approach to solving
the reliability problem \citep{Rubinstein2008}. Recasting Eq.(\ref{eq:002}) as:
\begin{equation}
  \label{eq:003}
   P_f= \int_{\Rr^M} \ve{1}_{\cd_f} (\ve{x}) \, f_{\ve{X}}(\ve{x}) \, d
   \ve{x} \equiv \Esp{\ve{1}_{\cd_f} (\ve{X})}
\end{equation}
the probability of failure is equal to the expectation of the
indicator function of the failure domain, which may be given the
following estimator using a $N$-sample $\mathfrak{X} = \acc{\Ve{X}^{(k)}
  , \; k=1 \enu N}$ made of $N$ independent copies of $X$:
\begin{equation}
  \label{eq:004}
  {\hat P_f} = \frac{1}{N}\sum\limits_{k = 1}^N 1_{\cd_f}\, \left(\ve{X}^{(k)}
  \right)  = \frac{N_f} {N}
\end{equation}
where $N_f$ is the number of samples that fall into the failure domain. This estimator is unbiased and mean-square convergent, since its
variance $\dsp{\Var{\hat P_f} = P_f(1-P_f)/N}$. The slow convergence
rate $\propto 1/\sqrt{N}$ makes the approach particularly
inefficient. It is easily shown that the coefficient of variation of the
estimator reads:
\begin{equation}
  \label{eq:005}
  CV_{P_f} \equiv \frac{\sqrt{\Var{\hat P_f}}}{\Esp{\hat P_f}}  
  \approx \frac{1}{\sqrt{N\,P_f}}
\end{equation}
From the above equation one can see that a typical evaluation of a
probability of failure of the order of magnitude $10^{-r}$ with
$CV_{P_f} \le 10\%$ requires about $10^{r+2}$ simulations. This number
is not affordable for low probabilities ($10^{-3} - 10^{-6}$) as soon as
the computational cost of each evaluation of $g$ (which includes a run
of $\cm$) is non negligible, \eg when finite element analysis is
involved.
 
\subsubsection{Beyond Monte Carlo simulation}
\label{sec:2.4.2}
Various methods have been proposed in the past 30 years in order to
solve the reliability problem efficiently. Broadly speaking they can be
classified as follows:
\begin{itemize}
\item methods that aim at decreasing the computational cost after
  introducing some approximation in the reliability estimation. The
  First Order (resp. Second Order) reliability methods (FORM/SORM) are
  well established in this area, see \citet{Hasofer1974, Rackwitz78,
    Breitung84, Hohenbichler1987, Breitung89, DK87, DK91} among others.
\item methods that are derived from Monte Carlo simulation with the goal
  of improving the convergence: {\em directional simulation}
  \citep{Ditlevsen87, Bjerager88}, {\em importance sampling}
  \citep{Hohenbichler1988, Melchers90, Maes1993, Au1999b} and more
  recently {\em subset simulation} \citep{Au2001,Katafygiotis2005,
    Hsu2010} and line sampling \citep{Pradlwarter2007}.
\item Methods that rely on the use of a {\em surrogate model}
  $\tilde{\cm}$ which is fast to evaluate and may be used {\em in place}
  of the original model $\cm$ for reliability analysis
\end{itemize}
A detailed presentation of the various classical methods can be found in
the textbooks by \citet{Ditlevsen1996, Melchers1999, Lemaire09}. This is the
aim of the present paper to address the last point and review different
classes of surrogate models and their use in structural reliability.

%%%%%%%%%%%%%%%%%%%%%%%%%%%%%%%%%%%%%%%%%%%%%%%%%%%%%%%%%%%%%%%%%%%%%%%%%%%%%%%%%%%%%%%%%%%%%%%%%
%\section{Meta-models for uncertainty propagation}

%%%%%%%%%%%%%%%%%%%%%%%%%%%%%%%%%%%%%%%%%%%%%%%%%%%%%%%%%%%%%%%%%%%%%%%%%%%%%%%%%%%%%%%%%%%%%%%%%
\section{Meta-models for structural reliability}
\label{sec:3}
\subsection{Introduction}
\label{sec:31}
As introduced above, a {\em meta-model} is an analytical
function with the following properties~:
\begin{itemize}
\item it belongs to a specific class of functions (the ``type'' of the
  meta-model) and it is fully characterized by a {\em set of parameters}
  once the class is selected;
\item it is fast to evaluate: carrying out some large size
  Monte Carlo simulation on the meta-model will be affordable;
\item it is fitted to the original model (also called ``true'' model in
  the sequel, \eg the limit state function $g$) using a set of
  ``observations'' of the true model, \ie a collection of input/output
  pairs (each observation is a computer experiment in the present context):
  \begin{equation}
    \label{eq:006}
    \mathfrak{X} = \acc{ \prt{\ve{x}^{(i)}, {y}^{(i)} = g
        \prt{\ve{x}^{(i)}}},  \; i =1 \enu N}
  \end{equation}
\end{itemize}

In the sequel, we will review the following classical types of
meta-models:
\begin{itemize}
\item linear (resp. quadratic) polynomial response surfaces
\item polynomial chaos expansions
\item Gaussian processes (also known as {\em kriging surrogates})
%\item Support vector machines
\end{itemize}

Note that {\em support vector machines} have been recently introduced in
the field of structural reliability by \citet{Hurtado2004a,
  Hurtado2004b}. Coming from the world of statistical learning this
technique is well adapted to the classification of a labelled
population, \eg a set of ``failure'' (resp. ``safe'') points in the
context of reliability. It has been recently combined with subset
simulation to provide a highly efficient way of assessing small
probabilities of failure \citep{Deheeger2007, DeheegerThesis,
  Bourinet2011}, see also \citet{Basudhar2008a, Basudhar2008b}. The
detailed presentation of this approach is beyond the scope of the
present paper.

\subsection{FOSM and FORM method viewed as linear response surfaces}
\label{sec:32}
The first-order second moment method (FOSM) \citep{Cornell1969} may be
interpreted as a type of linear response surface. Indeed Cornell's
reliability index is defined as the ratio between the mean value and
standard deviation of the safety margin defined by the performance
function $g$:
\begin{equation}
  \label{eq:007}
  \beta_{C} = \frac{\mu_g(\Ve{X})} {\sqrt{\Var{g(\Ve{X})}}}
\end{equation}
The latter variance is then obtained from a Taylor series expansion of
the performance function around the mean value of the input vector
denoted by $\ve{\mu}$, which reads:
\begin{equation}
  \label{eq:008}
  g(\ve{x}) = g(\ve{\mu}) + \nabla g \tr\prt{\ve{\mu}} 
\cdot \prt{\ve{x} - \ve{\mu}} 
+ o \prt{ \norme{\ve{x} -\ve{\mu}}{}^2}
\end{equation}
From the above equation one gets~:
\begin{equation}
  \label{eq:009}
  \Var{g(\Ve{X})} \approx \Esp{\bra{g(\ve{X}) - g(\ve{\mu})}^2} =
\nabla g\tr \prt{\ve{\mu}} \cdot \matcov \cdot \nabla g \prt{\ve{\mu}} 
\end{equation}
where:
\begin{equation}
  \label{eq:009b}
  \matcov = \Esp{\prt{\ve{X} -\mu}\prt{\ve{X} -\mu}\tr }
\end{equation}
is the covariance matrix of $\Ve{X} $. In other words, Cornell's
reliability index is implicitly based on a linear response surface built
up around the input parameters' mean values.

Due to a well-known lack of invariance when changing the performance
function \citep[Chap.~5]{Ditlevsen1996}, the famous {\em Hasofer-Lind}
reliability index $\beta_{HL}$ has been proposed \citep{Hasofer1974},
whose derivation may be summarized as follows in the context of the
so-called {\em first-order reliability method} (FORM):
\begin{itemize}
\item the input variables are transformed into a standard normal space
  by a suitable isoprobabilistic transform $\ct : \, \ve{x} \mapsto \ve{u}
  = \ct(\ve{x})$;
\item the reliability index $\beta_{HL}$ is defined as the algebraic
  distance between the origin of this space and the (transformed) limit
  state surface;
\item the associated probability of failure is obtained from the
  {\em linearization} of the limit state surface at the design point $\ve{u}^*$:
  \begin{equation}
    \label{eq:010}
    P_{f,{\text{FORM}}} = \Phi(-\beta_{HL})
  \end{equation}
where:
\begin{equation}
  \label{eq:011} \beta_{HL} = \text{sign}\, g\prt{\ct^{-1}(\ve{0})}
  \norme{\ve{u}^*}{}\qquad \ve{u}^* = \arg \underset{\ve{u} \in \Rr^M}
  {\min}\acc{ \norme{\ve{u}}{}^2 \; : \;
    g(\ct^{-1}(\ve{u})) \le 0}
  \end{equation}
\end{itemize}
In this respect again, the probability of failure is evaluated after
constructing a particular linear response surface, namely that obtained
by a Taylor series expansion of the transformed limit state (in the
standard normal space) around the design point $\ve{u}^*$.

\subsection{Quadratic response surfaces}
\label{sec:33}
Using a more classical setting, quadratic polynomial response surfaces may be
defined as follows:
\begin{equation}
  \label{eq:012}
  \tilde{g}(\ve{x}) = a_0 + \sum_{i=1}^M a_i \, x_i + \sum_{i=1}^M
  a_{ii} \, x_i^2 + \sum_{1\le i < j 
    \le M} a_{ij} \, x_i x_j
\end{equation}
which may be condensed as follows:
\begin{equation}
  \label{eq:013}
  \begin{split}
    \tilde{g}(\ve{x}) &=  \ve{f}(\ve{x})\tr \cdot \ve{a} \\
    \ve{a}\tr& = \prt{a_0,\, a_1 \enu a_M \,,\, a_{11}\enu 
      a_{MM} \,,\, \acc{ a_{ij} \;,
      \;1\le i< j\le M }}\\
    \ve{f}(\ve{x}) \tr&= \prt{1,\, x_1, \, \dots, x_M \, ,x_{1}^2,\, \dots,
      x_{M}^2 \,,\,  \acc{ x_i x_j \;, \; 1\le i< j\le M }}
  \end{split}
\end{equation}
In order to fit the response surface a set of observations $\mathfrak{X}
=\acc{ \prt{\ve{x}^{(i)}, g \prt{\ve{x}^{(i)}}}, \; i =1 \enu N}$ is
selected and the vector of coefficients $\ve{a}$ (of size $P$) is
computed by {\em minimizing the least-square error} between $g$ and
$\tilde{g}$:
\begin{equation}
  \label{eq:014}
  \ve{a}_{MS}= \arg \underset{\ve{a} \in \Rr^{P}} {\min} \sum_{i=1}^N 
  \bra{ g(\ve{x}^{(i)} ) -  \ve{f}\tr(\ve{x}^{(i)}) \cdot \ve{a}}^2
\end{equation}
The solution to this problem reads:
\begin{equation}
  \label{eq:015}
   \ve{a}_{MS}= \prt{\matF \tr \matF}^{-1} \matF \tr \ve{\Gamma}
\end{equation}
where $\matF$ is the information matrix of size $N \times P$  :
\begin{equation}
  \label{eq:016}
  \matF_{ij} = f_j (\ve{x}^{(i)})  \qquad \qquad  \ve{\Gamma} =
  \prt{g(\ve{x}^{(1)})  \enu g(\ve{x}^{(N)}) }\tr
\end{equation}
The use of quadratic polynomial response surfaces as a surrogate of the
limit state function has been pioneered by \citet{Faravelli1989}.
Various variants have been proposed throughout the 90's depending on the
choice of the polynomials (\eg taking the cross terms $x_i x_j$ into
account or not) and the experimental design used for the model fitting,
see \citet{Bucher1990, Rajashekhar1993, Kim1997, Das2000}. The use of
these approaches together with finite element models has been
popularized by \citet{Lemaire1998} and \citet{Pendola2000} where the
response surfaces are refined in an adaptive manner around the design
point obtained at each iteration (see also \citet{Gayton2003}). Further
applications may be found in \citet{Duprat2006} (concrete structures)
and \citet{Leira2005}, among others.

As a conclusion, quadratic response surfaces have been widely used in
the last 20~years for reliability analysis. However they lack of
versatility in the sense that a second order polynomial function can
only mimic models with smooth behaviours. Moreover it is implicitely
supposed that there is a single design point as in FORM, around which
the approximation may be built. This condition is seldom encountered in
industrial problems, which has lead to the use of more versatile
meta-models.

%Note that the various Second Order Reliability Methods (SORM)
%\citep{Breitung84, Breitung89, DK87, Hohenbichler1987, DK91} also r

\subsection{Polynomial chaos expansions}
\label{sec:34}
\subsubsection{Some history}
\label{sec:341}
Polynomial chaos (PC) expansions have been introduced in the literature
on stochastic mechanics in the early 90's by \citet{Ghanembook1991} and
have been limited to solving stochastic finite element problems
throughout the 90's. In the original setting, a boundary
value problem is considered in which some parameters are modelled by
{\em random fields}. The quantities of interest are the resulting
stochastic displacement and stress fields. Thus the use of PC expansions
has been intimately associated with spatial variability and considered
as a separate topic with respect to structural reliability for a while.

Considering the expansion in itself as a meta-model that is suitable for
reliability analysis has been originally explored by \citet{Sudret2000,
  Sudret2002, SudretIFIP2003}. Later on, the use of PC expansions has
blown up with the emergence of so-called {\em non intrusive} methods.
More specifically the regression approach has been developed and applied
to reliability analysis in \citet{BerveillerPMC04, Choi2004b,
  Berveiller2006a}, among others.

%Various applications of PC expansions to reliability analysis may be
%found in nuclear engineering \citep{SudretBlatmanIFIP2010}

\subsubsection{Polynomial chaos basis}
\label{sec:342}
Without going into too much mathematical details, one may consider
polynomial chaos expansions as an {\em intrinsic representation} of a
random variable that is defined as a function of the input random vector
$\Ve{X} $ . In the context of structural reliability the limit state
function leads to define the ``random margin'' $G = \cm(\Ve{X} ) $. The
probability of failure is then defined by $ P_f = \Prob{G \le 0}$.
Assuming that this variable has a finite variance and that the input
parameters in $\Ve{X} $ are independent (for the sake of simplicity in
this presentation \footnote{It is always possible to transform the
  original vector into independent variables using \eg the Nataf or
  Rosenblatt transform.}), the following representation holds
\citep{Soize2004}:
\begin{equation}
  \label{eq:018b}
 G =g(\Ve{X}) =\dsp{\sum_{\ua \in \Nn^M}  a_{\ua} \, \Psi_{\ua}(\Ve{X}) }
\end{equation}
In this equation the $\Psi_{\ua}(\Ve{X})$ are {\em multivariate
  orthonormal polynomials} in the input variables and $a_{\ua}$ are
coefficients to be computed. Since the component of $\Ve{X} $ are
independent, the joint PDF is the product of the margins. For each
marginal distribution $f_{X_i}(x_i)$ a functional inner product is
defined:
\begin{equation}
  \label{eq:019}
\langle \phi_1, \phi_2 \rangle_{i} \equiv \int_{\cd_i} \phi_1(x) \,
  \phi_2(x) \, f_{X_i}(x) \, dx
\end{equation}
For each variable $i=1 \enu M$ a family of polynomials is then built
which satisfies the following orthogonality properties:
\begin{equation}
\left\langle   {P_j^{(i)},P_k^{(i)}} \right\rangle = \int_{\cd_i} 
{P_j^{(i)}(x)\;P_k^{(i)}(x)\;{f_{{X_i}}}(x)\,dx = } \;a_j^i\;{\delta _{jk}}
\label{eq:020}
\end{equation}
where $\delta_{jk}$ is the Kronecker symbol which is equal to 1 if $j=k$
and 0 otherwise. The norm of polynomial $P_j^{(i)}$ is $a_j^i$ which is
usually not equal to 1. Thus in order to build an {\em orthonormal} family,
the above polynomials are rescaled. Classical families of polynomials
correspond to classical types of PDFs, namely Hermite polynomials are
orthogonal w.r.t to the Gaussian PDF, Legendre polynomials w.r.t to the
uniform PDF, etc. \citep{Xiu2002}, see Table~\ref{tab:01} for their
expressions and the associated normalization.

\begin{table} 
\tbl{Classical orthogonal polynomials}
{\tabcolsep24pt
\begin{tabular}{@{}p{2.5pc}p{4pc}p{5pc}p{5pc}@{}}\toprule 
 Distribution & PDF  & Orthogonal polynomials &
    Orthonormal basis \par$\psi_k(x)$ \\  \colrule
 Uniform & ${\mathbf 1}_{]-1,1[}(x) /2 $  & Legendre $P_k(x)$ &
    $P_k(x)/  \sqrt{\frac{1}{2k+1}}$ \\
    Gaussian & $ \frac{1}{\sqrt{2 \pi}} e^{-x^2/2}$ & Hermite
    $H_{e_k}(x)$ &  $H_{e_k}(x)/  \sqrt{k!}$\\
    Gamma& $x^a\, e^{- x} \,{\mathbf 1}_{\Rr^+} (x)$ &
    Laguerre $L^a_k(x)$ & $L^a_k(x)/\sqrt{\frac{\Gamma(k+a+1)}{k!}}$\\
    Beta & ${\mathbf 1}_{]-1,1[}(x) \, \frac{(1-x)^a(1+x)^b}{B(a)\,
      B(b)} $ & Jacobi $J^{a,b}_k(x)$ 
    &  $J^{a,b}_k(x)/{\mathfrak{J}}_{a,b,k}$ \\
    & & \multicolumn{2}{c}{
      $\mathfrak{J}_{a,b,k}^2 
      = \frac{2^{a+b+1}}{2k+a+b+1}
      \frac{\Gamma(k+a+1)\Gamma(k+b+1)}{\Gamma(k+a+b+1) \Gamma(k+1)}$}
    \\
\botrule
\end{tabular}}
\label{tab:01}
%\tabnote[{\it Source}]{Sample typeset by RPS for APSSRA12 conference proceedings.}
\end{table}

Once the univariate orthonormal polynomials are available, the
multivariate polynomials are built by tensorization. To each $M$-tuple
$\ua = \acc{\alpha_1 \enu \alpha_M } \in \Nn^M$ one associates the
polynomial $\Psi_{\ua}(\ve{x})$ as follows:
\begin{equation}
  \label{eq:021}
  \Psi_{\ua}(\ve{x}) = \prod_{i=1}^M \Psi_{\alpha_i}^{(i)} (x_i)
\end{equation}
The family of $\Psi_{\ua}$'s naturally inherits from the orthonormality of
univariate polynomials so that:
\begin{equation}
  \label{eq:022}
  \Esp{\Psi_{\ua} \, \Psi_{\ub}} = \delta_{\ua \ub}
\end{equation}
where $ \delta_{\ua \ub}$ is equal to 1 if the $M$-tuples $\ua$ and
$\ub$ are identical and zero otherwise.  Once the basis is built, a
troncature scheme has to be selected in order to carry out the
computation of the coefficients. The classical setting consists in
selecting {\em all the polynomials} of total degree $ |\ua| =
\dsp{\sum_{i=1}^M \alpha_i}$ not greater than a given $p$, \ie:
\begin{equation}
  \label{eq:023}
  g(\Ve{X}) \approx  g^{PC}(\Ve{X} ) \equiv \dsp{\sum_{\ua \in
      \ca}  a_{\ua} \, \Psi_{\ua}(\Ve{X}) } 
 \qquad  \text{where} \quad\ca =\acc{\ua \in \Nn^M\,:\, |\ua| \le p}
\end{equation}
This type of {\em a priori truncation} is somehow arbitrary although a
value of $p=2$ usually provides fair results for estimating the mean and
variance of the margin $G$ whereas $p=3$ is required in order to compute
probabilities of failure downto $10^{-4}$ with a satisfactory accuracy
\citep{SudretHDR}.  Note that error estimates have been recently
proposed together with adaptive algorithms in order to avoid the problem
of the {\em a priori} selection of $\ca$, as shown in
\citet{BlatmanPEM2010, SudretJCP2011}

\subsubsection{Computation of the coefficients}
\label{sec:343}
The original approach to computing the coefficients of a truncated PC
expansion in computational stochastic mechanics is of Galerkin-type
\citep{Ghanembook1991} and it is termed {\em intrusive} since it
requires the {\em ad-hoc} derivation of a weak formulation of the
underlying mechanical problem and its discretization.  {\em Non
  intrusive methods} have emerged in the literature as of 2002 and may
classified as follows:
\begin{itemize}
\item projection methods that make a direct use of the orthogonality
  properties of the PC basis:
  \begin{equation}
    \label{eq:024}
     a_{\ve{\alpha}} =  \Esp{G \, \Psi_{\ve{\alpha}}(\Ve{X})}  =\int_{{\cal D}_{\ve{X}}} g(\ve{x}) \,
  \Psi_{\ve{\alpha}}(\Ve{x}) \, f_{\ve{X} }(\ve{x}) \, d \ve{x} 
  \end{equation}
  The latter integral may be computed using either tensorized or sparse
  quadrature rules \citep{Ghiocel2002,Lemaitre02,Keese05}
\item stochastic collocation methods, which are based on Lagrange
  interpolation in the stochastic space and are essentially equivalent
  to the former \citep{Xiu:Hesthaven:2005, Xiu2009a};
\item regression methods which were introduced in the field of PC
  expansions in \citet{Berveiller2006a}
  % based on the early work by \citet{Isukapalli}
and recently improved through error estimation and
  adaptivity by \citet{BlatmanThesis,BlatmanPEM2010, SudretJCP2011}.
\end{itemize}

In this paper we will concentrate on the latter approach which is
similar in essence to building a quadratic polynomial response surface
as shown in Section~\ref{sec:33}, except that the basis functions are
now the members of the truncated PC expansion. The key idea consists in
considering the random margin $G$ as the sum of a truncated PC expansion
and a residual.
\begin{equation}
    \label{eq:025}
    G =g(\Ve{X} )= \sum_{\ua \in
      \ca}  a_{\ua} \, \Psi_{\ua}(\Ve{X})  + \vare
  \end{equation}
  The coefficients $a_{\ua}$ are obtained by minimizing the mean square
  residual:
\begin{equation}
    \label{eq:026}
    \ve{a}= \arg \underset{\ve{a} \in \Rr^{\text{card} \ca}} {\min}
    \Esp{\prt{g(\Ve{X}) - \sum_{\ua \in
          \ca}  a_{\ua} \, \Psi_{\ua}(\Ve{X})}^2}
\end{equation}
 which is approximated by using an experimental design of size $N$ as in
 Eq.(\ref{eq:014}):
\begin{equation}
    \label{eq:027}
    \ve{a}_{MS}= \arg \underset{\ve{a} \in \Rr^{\text{card} \ca}} {\min} \sum_{i=1}^N 
    \bra{ g(\ve{x}^{(i)}) - \sum_{\ua \in
        \ca}  a_{\ua} \, \Psi_{\ua}(\ve{x}^{(i)})}^2
\end{equation}
This leads to solving a linear system as in Eq.(\ref{eq:015}). In
contrast to the projection and the stochastic collocation methods which
make use of sparse grids, the experimental design is selected here so as
to be {\em space-filling}: Latin Hypercube sampling \citep{McKay79} or
quasi-random numbers are used for this purpose. Empirically the size of
the experimental design is selected as follows: $N$=2-3~Card$\ca$
\citep{BlatmanThesis}.

\subsubsection{Adaptive PC expansions and application to structural
  reliability}
\label{sec:344}
Once the basis is built and the coefficients have been computed the PC
expansion is treated as a global polynomial response surface and
substituted for the ``true'' limit state function for computing the
probability of failure:
\begin{equation}
  \label{eq:028}
  P_f \approx   P_f^{PC} \equiv \Prob{G^{PC} \le 0} =  \int_{\acc{\ve{x} 
      \,:\, g^{PC}(\ve{x}) \le 0}} f_{\ve{X}}(\ve{x}) \, d \ve{x}  
\end{equation}
As $ g^{PC}(\ve{x}) = \dsp{\sum_{\ua \in \ca} a_{\ua} \,
  \Psi_{\ua}(\Ve{X}) }$ is polynomial and straightforward to evaluate,
crude Monte Carlo simulation may be used to compute $ P_f^{PC}$,
although more advanced simulation techniques such as subset simulation
may be used.

The main difference between this approach and the traditional quadratic
response surfaces are:
\begin{itemize}
\item an arbitrary high degree of polynomials may be used, which allows
  one to fit complex limit state functions beyond a quadratic
  approximation.
\item error estimates based on cross-validation techniques
  \citep{Stone1974} are available \citep{BlatmanPEM2010} which may be
  coupled with adaptive algorithms that automatically detect the best
  sparse PC representation \citep{SudretJCP2011} and increase the
  maximal degree of the PC expansion as long as the prescribed
  admissible error is not attained.
\item in order to avoid {\em overfitting} when computing the
  coefficients, the size of the experimental design $N$ is increased
  automatically so as to always satisfy the condition $N \le 2-3\,
  \text{Card}\, \ca$.
\item by construction the PC expansion readily provides useful
  information on the moments of the performance $G^{PC} = g^{PC}(
  \Ve{X})$ as well as useful information on sensitivity analysis
  \citep{SudretRESS2008b}.
\end{itemize}
A synthesis of such an approach with applications in structural
reliability may be found in \citet{SudretWiley2011b}.

%%%%%%%%%%%%%%%%%%%%%%%%%%%%%%%%
\subsection{Kriging meta-models}
\label{sec:35}
\subsubsection{Introduction}
\label{sec:351}
Building surrogate models in order to reduce the computational cost
associated with reliability analysis, stochastic or optimization
problems has been given the generic name of {\em computer experiments}
in the statistical literature.  In this respect {\em kriging} has
emerged in the last two decades as a powerful tool for building
meta-models. It has not been used in structural reliability until
recently though.

Historically kriging was named after the South African engineer D.~Krige
who initiated a statistical method for evaluating the mineral ressources
and reserves \citep{Krige1951}. This opened the field of {\em
  geostatistics} later formalized by \citet{Matheron:1963}, see also
\citet{Cressie1993, Chiles:Delfiner:1999}. The term {\em kriging} has
been coined in order to honor the seminal work of D.~Krige. The basic
idea is to model some function known only at a finite number of sampling
points as the realization of a Gaussian random field. In this setting
the sampling space is a ``physical'' two- or three-dimensional space.

Later, \citet{Sacks1989} introduced the key idea that kriging may also
be used in the analysis of computer experiments in which:
\begin{itemize}
\item the data is not measured but results from evaluating a computer
  code, \ie a simulator such as a finite element code;
\item the points where data is collected are not physical coordinates in
  a 2D or 3D space, but parameters in an abstract space of arbitrary
  size $M$.
\end{itemize}
In contrast to polynomial chaos expansions kriging provides a meta-model
that does not depend on the probabilistic model for the input random
vector $\Ve{X} $.

%%%
\subsubsection{Mathematical setting}
\label{sec:352}
The modern setting of kriging for computer experiments (also called
Gaussian process modelling \citep{Santner2003}) reads as
follows. The function-to-surrogate (\eg the limit state function
$g$ in the context of structural reliability) is supposed to be
a realization of a Gaussian process denoted by $Y(\ve{x}, \omega)$
defined as follows:
\begin{equation} 
   \label{eq:029}
  Y\left(\ve{x} , \, \omega\right) =
  \ve{f}\left(\ve{x}\right)\tr\,\ve{a} + Z\left(\ve{x}, \omega\right)
\end{equation} 
In this equation $\ve{f}\left(\ve{x}\right)\tr\,\ve{a}$ is the mean of
the process, which is represented by a set of basis functions
$\acc{f_i,\;i = 1\enu P} $ (\eg polynomial functions) and
$Z\left(\ve{x}, \omega\right)$ is a stationary zero mean Gaussian
process with variance $\sigma_Y^2$ and autocorrelation
function\footnote{For the sake of clarity the notation $\omega$ that
  recalls the randomness of the various quantities is abandoned in the
  sequel.}:
 \begin{equation}
   \label{eq:030}
   C_{YY}\left(\ve{x},\,\ve{x}'\right) =
   \sigma_Y^2\,R\left(\ve{x}-\ve{x}'\, ,\,\ve{\theta}\right), \quad
   \left(\ve{x},\,\ve{x}'\right) \in \cd_{\ve{X}}\times\cd_{\ve{X}}
 \end{equation}
 In the above equation $\ve{\theta}$ gathers all the parameters defining
 $C_{YY}$. In practice, square exponential models are generally
 postulated: 
\begin{equation}
  \label{eq:031}
  R\left(\ve{x}-\ve{x}',\ve{\theta}\right) =
  \exp\left(\sum\limits_{k=1}^M -
    \left(\frac{x_k-x_k'}{\theta_k}\right)^2\right)
\end{equation}
although other types of autocorrelation models such as generalized
exponentials or the Mat\'ern kernel may be used
\citep{Santner2003}.

In order to establish the kriging surrogate a set of computer
experiments is run and gathered in a vector $ \ve{\Gamma} =
\prt{g(\ve{x}^{(1)}) \enu g(\ve{x}^{(N)}) }\tr$. The kriging estimator
at a given point $\ve{x} \in \cd_{\ve{X}}$ is by definition a Gaussian
random variate $\widehat{Y}\left(\ve{x}\right) \sim
\cn\left(\mu_{\widehat{Y}} \left(\ve{x}\right),\,
  \sigma_{\widehat{Y}}\left (\ve{x}\right)\right)$ obtained by requiring
that it is the {\em best linear unbiased estimator} (BLUE) of
$g(\ve{x})$ conditioned to the observations gathered in $\mathfrak{X} $.
In other words it is obtained as a linear combination of the
observations and it is unbiased with minimum variance. After some rather
lengthy algebra the kriging estimator reads as follows (see
\citet{Santner2003}, \citet[Chap.~1]{DubourgThesis}):
\renewcommand{\ma}[1]{\textbf{\textrm{#1}}}
\begin{equation}\label{eq:032}
    \mu_{\widehat{Y}}\left(\ve{x}\right) =
    \ve{f}\left(\ve{x}\right)\tr\,\hat{\ve{a}} +
    \ve{r}\left(\ve{x}\right)\tr{\ma{R}}^{-1}\left(\ve{\Gamma} -
    \ma{F}\,\hat{\ve{a}}\right)
\end{equation}
In this equation the following notation $\ve{r}$, $\ma{R}$ et $\ma{F}$
is used:
\begin{eqnarray}
    r_i(\ve{x}) & = &
    R\left(\ve{x}-\ve{x}^{(i)},\,\ve{\theta}\right),\;i=1\enu N
%    \nolabel
\\[-5pt] 
    \ma{R}_{ij} & = &
    R\left(\ve{x}^{(i)}-\ve{x}^{(j)},\,\ve{\theta}\right),\;i=1\enu
    N,\;j=1\enu N  \label{eq:034} \\[-5pt] 
    \ma{F}_{ij} & = & f_j\left(\ve{x}^{(i)}\right),\;i=1\enu p,\;j=1\enu N  
%\nolabel
\end{eqnarray}
On top of the {\em mean prediction} given in Eq.(\ref{eq:032}), the
kriging approach yields the so-called {\em kriging variance}
$\sigma_{\widehat{Y}}\left(\ve{x}\right)$ which corresponds to an
epistemic uncertainty of prediction that is related to the finite size
of the available observation data gathered in $\mathfrak{X}$. This
variance reads:
\begin{eqnarray} 
  \label{eq:035}
  \sigma_{\widehat{Y}}^2\left(\ve{x}\right) & = & \sigma_{Y}^2\,\left(1 -
    \left\langle\begin{array}{cc}
        \ve{f}\left(\ve{x}\right)\tr & \ve{r}\left(\ve{x}\right)\tr
      \end{array}\right\rangle\,
    \left[\begin{array}{cc}
        \ma{0} & \ma{F}\tr \\
        \ma{F} & \ma{R}
    \end{array}\right]^{-1}\,
  \left[\begin{array}{c}
      \ve{f}\left(\ve{x}\right)\\
      \ve{r}\left(\ve{x}\right)
    \end{array}\right]
\right)
\end{eqnarray}

So far the ``regression'' part of the kriging estimator $
\ve{f}\left(\ve{x}\right)\tr\,\hat{\ve{a}}$ in Eq.(\ref{eq:032}) and the
parameters $\ve{\theta}$ in Eq.(\ref{eq:031}) have not been solved for.
The classical approach consists in building an {\em empirical BLUE} by
using a likelihood function from the joint (Gaussian) distribution of
the observations. Indeed by the underlying assumption of kriging, the
values gathered in $\ve{\Gamma}$ form a single realization of a
Gaussian vector $\acc{Y^{(1)} \enu Y^{(N)} }$. Then the likelihood of
the observations in $\ve{\Gamma}$ is maximized with respect to
$\acc{\ve{a}, \sigma_Y^2, \ve{\theta}}$. It may be shown that this
optimization yields an analytical expression for $\acc{\ve{a},
  \sigma_Y^2}$ as a function of $\ve{\theta}$:
\begin{eqnarray}
  \label{eq:036}
  \ve{a} &=&
  \left(\ma{F}\tr\,\ma{R}^{-1}\,\ma{F}\right)^{-1}
  \,\ma{F}\tr\,\ma{R}^{-1}\,\ve{\Gamma}   \\\label{eq:036b}
  \sigma_Y^2 &=& \frac{1}{N}\left(\ve{\Gamma} -
    \ma{F}\,\ve{a}\right)\tr \,{\ma{R}}^{-1}
  \,\left(\ve{\Gamma} -    \ma{F} \,\ve{a}\right)
\end{eqnarray}
In these expressions the dependence $\ma{R}(\ve{\theta})$ (see
Eq.(\ref{eq:031})) has been omitted for the sake of clarity.  The best
fit values of $\ve{\theta}$ are eventually obtained from a numerical
optimization, see \eg \citet{Marrel2008, DubourgThesis} for details.

The great features of kriging compared to the polynomial response
surfaces presented in Sections~\ref{sec:33}-\ref{sec:34} are summarized
below:
\begin{itemize}
\item The mean kriging estimator given in Eq.(\ref{eq:032}) is
  interpolating the data, meaning that
  $\mu_{\widehat{Y}}\left(\ve{x}^{(i)}\right) =
  g\left(\ve{x}^{(i)}\right)$ and the kriging variance is zero in these
  points: $\sigma^2_{\widehat{Y}}\left(\ve{x}^{(i)}\right) = 0$
\item The kriging variance $\sigma^2_{\widehat{Y}}(\ve{x})$ is interpreted
  as a measure of the epistemic uncertainty of prediction in each point
  $\ve{x} $. It shall not to be confused with the aleatoric uncertainty
  represented through random vector $\Ve{X} $ that is related to the
  probability measure $\Pp\bra{d \ve{x} } = f_{\Ve{X} }(\ve{x} ) \, d
  \ve{x} $.  Thus $\sigma^2_{\widehat{Y}}(\ve{x})$ may be used as an
  indicator for adaptively enrich the experimental design and refine the
  meta-model.
\end{itemize}

%%%
\subsubsection{Adaptive kriging and applications to structural reliability}
\label{sec:353}
Kriging has been first used for structural reliability problems in the
contributions by \citet{Romero2004} and \citet{Kaymaz2005}. In these
early papers the experimental design is either fixed or enriched
``passively'', \ie not taking into account the information brought by
the kriging variance.

Enriching sequentially the experimental design by using a criterion that
leads to adding points in the vicinity of the limit state function has
been proposed by \citet{Bichon2008} under the acronym EGRA (efficient
global reliability analysis). The authors define an {\em expected
  feasibility function} $EFF(\ve{x})$ that provides an indication on the
vicinity of the current point to the true limit state function. Starting
from the premise that only points close to the limit state function
bring additional information to build the surrogate, they maximize this
criterion at each step to get the next point to be added to the
experimental design for the next iteration.

Recently a similar approach called AK-MCS (for ``active kriging + Monte
Carlo simulation'') has been devised by \citet{Echard2011}, who
propose an epistemic error function which is directly based on the
kriging variance. The so-called $U$-function is defined by:
\begin{equation}
  \label{eq:037}
  U(\ve{x}) = \frac{
   | \mu_{\widehat{Y}}\left(\ve{x}\right)|
 }
  {\sigma_{\widehat{Y}}\left(\ve{x}\right)}  
\end{equation}
A small value of $U(\ve{x})$ means that either the kriging-limit state
function (kriging-LSF) $\mu_{\widehat{Y}}\left(\ve{x}\right)$ is close
to zero (vicinity of $\ve{x} $ to the surrogate limit state function )
or $\sigma_{\widehat{Y}}\left(\ve{x}\right)$ is large (large uncertainty
prediction) or both.  Once a first kriging prediction has been carried
out using a small size experimental design, this $U$-function is
evaluated on a large-size Monte Carlo sample. Note that it does not
require new calls to the $g$-function. The point leading to the smallest
value of $U(\ve{x})$ is added to the experimental design for improving
the kriging predictor at the next iteration.

In both approaches, the active enrichment of the experimental design is
carried out {\em sequentially}, \ie one single point is added from one
iteration to the other. This may be considered as a weakness since
usually the criteria to optimize show several local extrema (in other
words, there are several candidate points that could be equally added at
each iteration). Moreover it is nowadays common to have distributed
computing facilities which enables parallel evaluations of the
limit state function for a set of values of $\ve{x}$.

\begin{figure}[ht]
  \centering
  \includegraphics[width=0.9\textwidth]{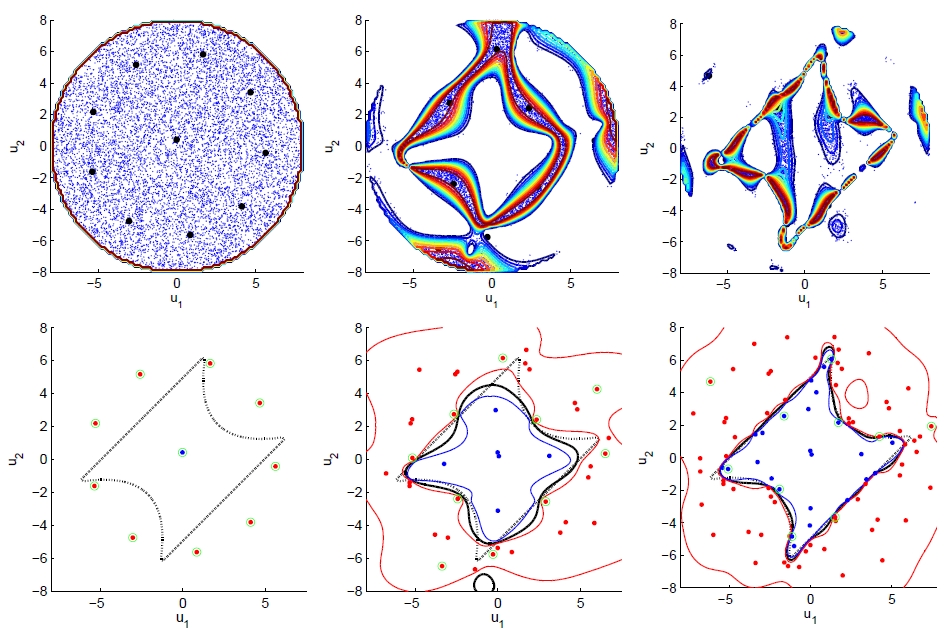}
  \caption{Illustration of the adaptive kriging technique (after
    \citet{SudretDubourgIcasp2011}). The original sample set is
    space-filling. In the later iterations, the sample points fall
    mainly within the margin of uncertainty defined by the kriging
    variance. Clustering is used in each step in order to select a few
    points to enrich the experimental design.}
  \label{fig:01}
\end{figure}

Starting from this observation \citet{DubourgThesis, Dubourg2011} have
proposed to define the following {\em probabilistic classification
  function}:
\begin{equation}
  \label{eq:038}
  \pi(\ve{x}, t) = \cp\bra{\hat Y(\ve{x}) \le t } =
    \Phi\prt{\frac{t-\mu_{\widehat{Y}}\left
          (\ve{x}\right)}{\sigma_{\widehat{Y}}\left(\ve{x}\right)}} 
\end{equation}
In this equation again, $\cp\bra{\bullet}$ denotes the Gaussian
probability measure associated with the epistemic uncertainty of kriging
and {\em not} the aleatoric uncertainty in $\Ve{X} $ (probability
measure denoted by $\Pp\bra{\bullet}$). The vicinity of any point
$\ve{x} $ to the kriging-LSF defined by $\acc{\ve{\xi} :
  \mu_{\widehat{Y}}(\ve{\xi}) = 0}$ is measured by $ \pi(\ve{x}, 0)$,
which shows some similarity with the $U$-function in Eq.(\ref{eq:037}).
From this function a {\em margin of uncertainty} $\mathfrak{M}$ is
defined which corresponds to ``confidence intervals'' around the
kriging-LSF:
\begin{equation}
  \label{eq:039}
  \mathfrak{M} = \acc{\ve{x}\,:\,-k\,\sigma_{{\widehat{Y}}}
    \left(\ve{x}\right) \leq \mu_{\widehat{Y}}\left(\ve{x}\right) \leq
    +k\,\sigma_{{\widehat{Y}}}\left(\ve{x}\right)} 
\end{equation}
where $k$ is a ``number of standard deviations'', \eg $k=1.96$ for a
95\% confidence interval. The enrichement criterion is then defined as
the probability of being in the margin of uncertainty at point $\ve{x}
$ which turns out to be:
\begin{equation}
  \label{eq:040}
  \cc({\ve{x} }) = \cp\bra{\widehat{Y}(\ve{x})\in \mathfrak{M}} = \pi(\ve{x},
  k\,\sigma_{\widehat{Y}}(\ve{x}) ) - \pi(\ve{x},
  -k\,\sigma_{\widehat{Y}}(\ve{x}) )
\end{equation}
This quantity $\cc(\ve{x} )$ is then considered as a probability density
function (up to a constant) from which it is simulated using a Markov
chain Monte Carlo algorithm such as the {\em slice sampling}
\citep{Neal2003}. Instead of picking up a {\em single} point as in
AK-MCS or EGRA, this approach yields a large set of points which
concentrate in the margin (see Figure~\ref{fig:01}).

A limited number $K$ of new experimental points (to be added to
$\mathfrak{X}$) is then obtained by clustering the large sample set and
added to the experimental design. Convergence criteria are established
based on the bounds on $P_f$ obtained by the boundaries of the margin
$\mathfrak{M}$, see \citet{Dubourg2011} for details.

\subsection{Conclusion}
\label{sec:36}
Quadratic response surfaces, polynomial chaos expansions and kriging
techniques have been reviewed as meta-modelling techniques for
structural reliability analysis. Whatever their respective advantages
and drawbacks, all these meta-models are usually used as {\em
  substitutes} for the original limit state function, meaning that the
probability of failure is simply evaluated as in Eq.(\ref{eq:028}). In
principle there is no guarantee that the probability of failure
evaluated on the surrogate is equal or sufficiently close to the that
obtained from the surrogate.  A way of addressing this issue is now
proposed.

%%%%%%%%%%%%%%%%%%%%%%%%%%%%%%%%%%%%%%%%%%%%%%%%%%%%%%%%%%%%%%%%%%%%%%%%%%%%%%%%%%%%%%%%%%%%%%%%%
\section{Meta-models as a means for importance sampling}
\label{sec:4}
\subsection{Introduction}
\label{sec:41}
As shown in the previous sections meta-models are usually built in a
first step of the reliability analysis and then substituted for the
``true'' limit state function. Even if some adaptive scheme allows one
to ensure the closeness of the original function and its surrogate, the
unbiasedness of the estimator of $P_f$ based on the surrogate cannot be
guaranteed.

In order to solve this problem, \citet{DubourgPEM2011} propose to
combine the well-known importance sampling technique with kriging in
order to get unbiased estimates of the probability of failure. As shown
below, the surrogate is used to build a ``smart'' importance sampling
density instead of being substituted as in Eq.(\ref{eq:028}). 

\subsection{Reminder on importance sampling}
\label{sec:42}
The reason why crude Monte Carlo simulation is inefficient for
evaluating probabilities of failure is that a typical sample will
contain a majority of points located in the central part of the input
distributions whereas the realizations that lead to failure are in the
tails. The key idea of importance sampling is to use an {\em
  instrumental density} which allows one to concentrate the drawn
samples in the region of interest.  Consider a non zero probability
density function $h(\ve{x} )$ defined on $\Rr^M$. Eq.(\ref{eq:003}) may
be recast as:
%\begin{equation}
%  \label{eq:043}
%  \forall \, \ve{x} \in \cd_{\Ve{X}} :\quad h(\ve{x} ) = 0 \Longrightarrow
%  \ve{1}_{\cd_f} (\ve{x}) \, f_{\ve{X}}(\ve{x}) =0
%\end{equation}
\begin{equation}
  \label{eq:044}
  \begin{split}
      P_f&= \int_{\Rr^M} \ve{1}_{\cd_f} (\ve{x}) \, \frac{
        f_{\ve{X}}(\ve{x})}{h(\ve{x} ) }  h(\ve{x} ) \, d
   \ve{x} \\
   &= \Espe{h}{\ve{1}_{\cd_f} (\ve{X}) \frac{
        f_{\ve{X}}(\ve{X})}{h(\ve{X} ) } }
  \end{split}
\end{equation}
where the subscript in $\Espe{h}{\bullet}$ means that the expectation is
taken with respect to the probability measure associated with $h$.  The
classical importance sampling (IS) estimator is built from a sample set
drawn from the instrumental distribution, say $\mathfrak{X}_h =
\acc{\ve{x}_h^{(i)} , \; i =1 \enu N}$:
\begin{equation}
  \label{eq:045}
  \hat{P}_{f,IS} \equiv \frac{1}{N} \sum_{i=1}^N  \ve{1}_{\cd_f}
  (\ve{x}^{(i)}_h) \frac{f_{\Ve{X} }(\ve{x}^{(i)}_h)} {h(\ve{x}^{(i)}_h)}
\end{equation}
The art of importance sampling consists in using an instrumental density
that minimizes the variance of the latter estimator.
\citet{Rubinstein2008} shows that the optimal density (which actually
reduces the variance of the estimator to 0) reads:
\begin{equation}
  \label{eq:046}
  h^*(\ve{x}) = \frac{\ve{1}_{\cd_f} (\ve{X}) \, f_{\Ve{x} }(\ve{x})}{
    \int_{\cd_{\Ve{x}}}\ve{1}_{\cd_f} (\ve{X}) \, f_{\Ve{X} }(\ve{x}) }
  =  \frac{\ve{1}_{\cd_f} (\ve{x}) \, f_{\Ve{X} }(\ve{x})} {P_f}
\end{equation}
However this density cannot be used in practice since it depends on the
unknown quantity of interest $P_f$. The key idea of {\em
  meta-model-based importance sampling} (called meta-IS for short) is to
surrogate this optimal instrumental density by kriging.

\subsection{Sub-optimal instrumental density}
\label{sec:43}
\begin{figure}[ht]
  \centering
 \includegraphics[width=0.9\textwidth]{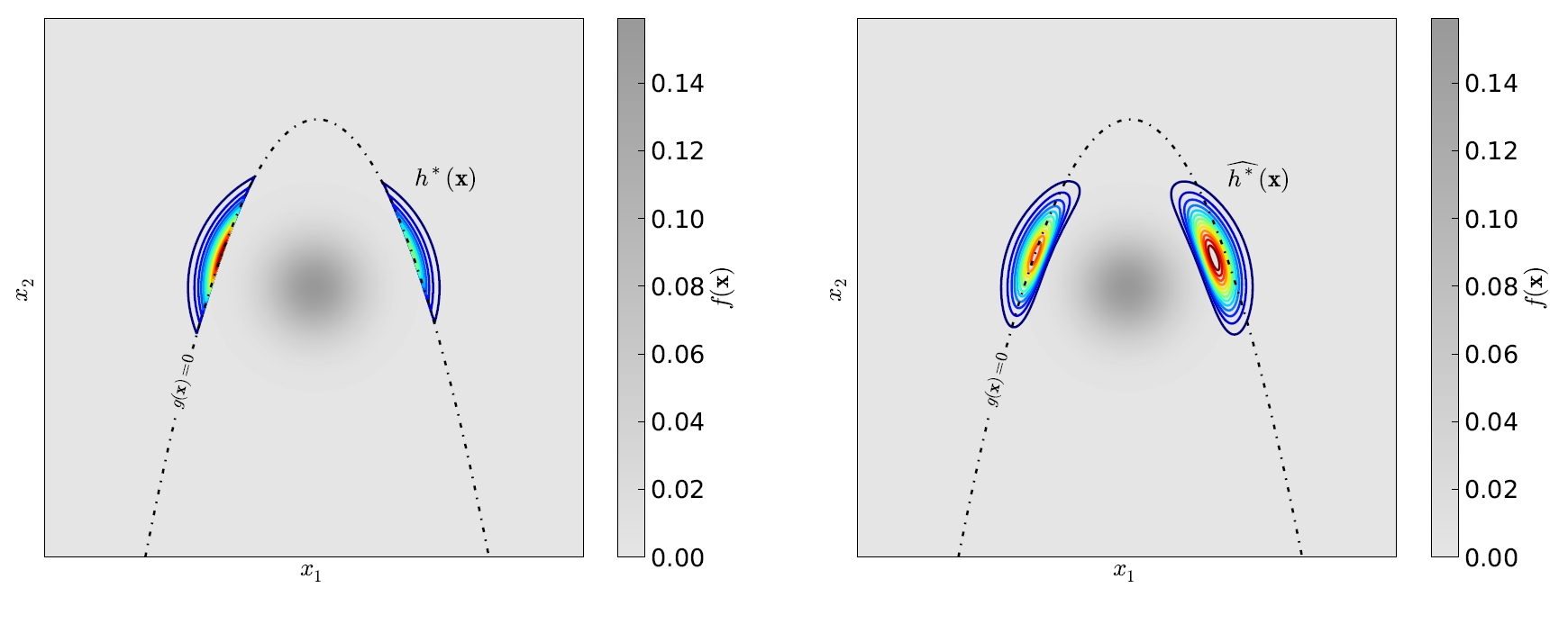}
 \caption{Comparison of the optimal (left) and the meta-IS (right)
   instrumental probability density functions (after
   \citet{DubourgPEM2011})}
  \label{fig:02}
\end{figure}

Assuming that some kriging meta-model $\widehat{Y}$ of the limit state
function is available, the indicator function of the failure domain may
be replaced by the probabilistic classification function (see
Eq.(\ref{eq:038})) evaluated for $t=0$:
\begin{equation}
  \label{eq:047}
  \pi(\ve{x} ) \equiv \Phi\prt{\frac{0-\mu_{\widehat{Y}}\left 
        (\ve{x}\right)}{\sigma_{\widehat{Y}}\left(\ve{x}\right)}} 
\end{equation}
Indeed if the surrogate is accurate in a given point $\ve{x}_0 $ then
$\sigma_{\widehat{Y}}(\ve{x}_0 )$ is close to zero and
$\mu_{\widehat{Y}}(\ve{x}_0)$ is close to $g(\ve{x}_0 )$. If the latter
is negative then $\mu_{\widehat{Y}}(\ve{x}_0)
/\sigma_{\widehat{Y}}(\ve{x}_0 ) \rightarrow -\infty $ and
$\pi(\ve{x}_0) \approx 0$. In contrast, if $g(\ve{x}_0 ) >0 $ then
$\mu_{\widehat{Y}}(\ve{x}_0) /\sigma_{\widehat{Y}}(\ve{x}_0 )
\rightarrow +\infty $ and $\pi(\ve{x}_0) \approx 1$. Thus $\pi(\ve{x} )$
is a kind of ``smoothed'' version of the indicator function
$\ve{1}_{\cd_f}$. This leads to the {\em meta-IS instrumental
density} (to be compared with Eq.(\ref{eq:046})) : 
\begin{equation}
  \label{eq:048}
  \tilde{h}(\ve{x}) \equiv \frac{\pi(\ve{x}) \, f_{\Ve{X}}(\ve{x})}{P_{f\vare}} =
  \frac{\Phi\prt{- \mu_{\widehat{Y}}(\ve{x})/ \sigma_{\widehat{Y}} 
      (\ve{x})}   \, f_{\Ve{X}}(\ve{x})}{P_{f\vare}}
\end{equation}
where the normalization constant $P_{f\vare}$ reads:
\begin{equation}
  \label{eq:049}
  P_{f\vare} \equiv \int_{\cd_{\Ve{X}}}
  \Phi\prt{- \mu_{\widehat{Y}}(\ve{x})/ \sigma_{\widehat{Y}} 
      (\ve{x})}   \, f_{\Ve{X}} (\ve{x}) \, d \ve{x} 
\end{equation}
As an example the optimal instrumental density and that obtained from the
above procedure is shown in Figure~\ref{fig:02} where the quadratic limit
state is taken from \citet{DerKiureghian1998}. 

\subsection{Meta-IS estimator of the probability of failure } 
\label{sec:44}
Substituting for Eq.(\ref{eq:048}) in Eq.(\ref{eq:044}) and using
Eq.(\ref{eq:049}) one gets:
\begin{equation}
  \label{eq:050}
  P_f = \alpha_{corr} \, P_{f\vare}
\end{equation}
where:
\begin{equation}
  \label{eq:051}
  \alpha_{corr}  = \Espe{\tilde{h}}
  {\frac{\ve{1}_{\cd_f} (\ve{X})}{\pi(\Ve{X} )}}
\end{equation}
is a correction factor that quantifies the error that has been made by
substituting the probabilistic classification function $\pi(\ve{x})$ to
the indicator function $\ve{1}_{\cd_f} (\ve{X})$.  The Monte Carlo
estimate of the latter then reads:
\begin{equation}
  \label{eq:052}
  \begin{split}
    \widehat{P_f}_{metaIS} &= \widehat{\alpha}_{corr} \,
    \widehat{P_{f\vare}} \\
    \text{where:} \qquad \widehat{\alpha}_{corr} & = {\frac{1}{N_{corr}}
      \sum_{k=1}^{N_{corr}} \frac{\ve{1}_{\cd_f} (\tilde{\ve{x}}^{(k)})}
      {\pi(\tilde{\ve{x}}^{(k)})} } \\
    \widehat{P_{f\vare}} &= {\frac{1}{N_\vare} \sum_{l=1}^{N_\vare}
      \pi(\ve{x} ^{(l)})}
  \end{split}
\end{equation}
In the above equation, the first sum is computed using $N_{corr}$
realizations following the meta-IS PDF $\tilde{h}$ while the second sum
is computed using $N_{\vare}$ realizations of $\Ve{X} $.

The estimator in Eq.(\ref{eq:052}) is proven to be unbiased and its
variance may be derived straightforwardly \citet{DubourgPEM2011}. From a
computational point of view, it is important to note that
$\widehat{P_{f\vare}} $ is evaluated at low cost since it only requires
the evaluation of the probabilistic classification function
Eq.(\ref{eq:047}) which is analytical when the kriging surrogate is
available. Thus $N_\vare = 10^{5-6}$ is affordable, leading to an almost
exact value of $P_{f\vare}$. In contrast, computing
$\widehat{\alpha}_{corr}$ requires the evaluation the ``true'' indicator
function of the failure domain. Thus $N_{corr}$ shall be limited to a
few hundredw in practice.

\subsection{Meta-IS estimator of the probability of failure } 
\label{sec:45}
As a conclusion, using meta-models as a tool for defining a
quasi-optimal instrumental density for importance sampling solves the
problem of possible biasedness resulting from a direct substitution.
This is a great advantage of meta-IS compared to the approaches by
\citet{Bichon2008,Echard2011}.

In order to optimize the efficiency a trade-off between a) the number of
calls to the $g$-function (say, $N$) required for building the
meta-model $\widehat{Y}$ and b) the number of samples $N_{corr}$ shall
be found, since the ``total cost'' is $N+N_{corr}$. A large $N$ will
provide a very accurate meta-model that will lead to $\alpha_{corr}
\approx 1$ at a high computational cost though. In contrast a ``medium
accurate'' surrogate will be compensated for by a correction factor
$\alpha_{corr} \ne 1$. In order to balance these two aspects a procedure
based on cross-validation is proposed by \citet{DubourgPEM2011} for
stopping the iterative improvement of the kriging surrogate
(Section~\ref{sec:353}) when $P_{f\vare}$ is sufficiently close to
$P_f$.

%%%%%%%%%%%%%%%%%%%%%%%%%%%%%%%%%%%%%%%%%%%%%%%%%%%%%%%%%%%%%%%%%%%%%%%%%%%%%%%%%%%%%%%%%%%%%%%%%
\section{Application examples}
\label{sec:5}
\subsection{Frame structure - Polynomial chaos expansion \citep{BlatmanPEM2010}}
\label{sec:51}
\begin{figure}[!ht]
  \begin{center}
    \includegraphics[trim = 10mm 10mm 10mm 150mm, clip, width=0.8\textwidth]{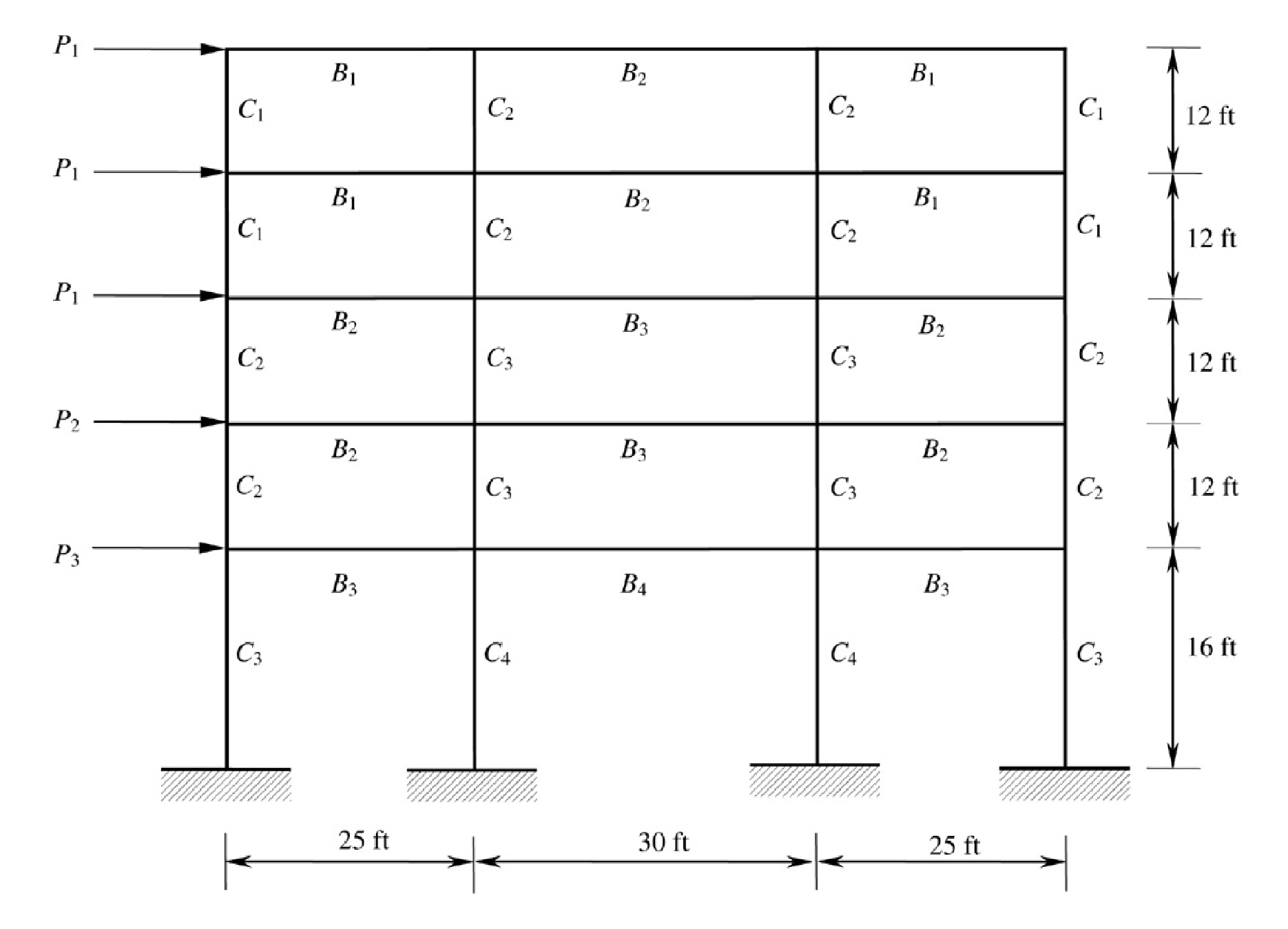}
    \caption{Sketch of a 3-span, 5-story frame structure subjected to
      lateral loads}
    \label{fig:Frame_Fig}
  \end{center}
\end{figure}

Let us consider a 3-span, 5-storey frame structure as the one sketched
in Figure~\ref{fig:Frame_Fig}. Of interest is the top floor displacement
when the structure is submitted to lateral loads. The associated
serviceability limit state function reads: 
\begin{equation}
  g(\ve{X}) \, = \, u_{max} - \text{\textsc{FEModel}}(\ve{X}) 
\end{equation}
where $u_{max}$ is a given threshold, \textsc{FEModel}$(\bullet)$ is the
finite element model (considered as a black box) that evaluates the top
floor horizontal displacement, and $\Ve{X} $ gathers 21~correlated
random variables which describe the uncertainty in the member's cross
section, inertia and applied loads, see \citet{BlatmanPEM2010} for the
complete description.  The probability of failure is investigated for
different values of the threshold, namely $u_{max} = 4- 9$~cm.
Reference values are obtained by FORM followed by importance sampling
($500,000$ model evaluations are used to get a coefficient of variation
less than $1.0\%$ on $P_f$).

Estimates of the reliability index are computed by post-processing a
full third-order PC expansion as well as a sparse PC approximation. The
latter is built up by setting the target approximation error equal to
$10^{-3}$ (this is an overall mean square error that drives the
convergence of the adaptive PC expansion, see the original paper for
details). The estimates of the various generalized reliability indices
$\beta = - \Phi^{-1}(P_{f})$ are reported in
Table~\ref{tab:Frame:Reliab}.

\begin{table}[!ht] 
  \tbl{Frame structure - Estimates of the generalized reliability
    index $\beta = - \Phi^{-1}(P_{f})$ for various values of the
    threshold displacement}
  {\tabcolsep8pt
\begin{tabular}{cccccc}
  \toprule
  ~~Threshold (cm)~~ & ~~Reference~~ & \multicolumn{2}{c}{Full PCE}& \multicolumn{ 2}{c}{Sparse PCE } \\
  &  $\beta^{REF}$ & $\widehat{\beta}$ & $\epsilon \; (\%)$ & $\widehat{\beta}$ & $\epsilon \; (\%)$ \\
  \colrule
  4 & 2.27 & 2.26 & 0.4 & 2.29 & 0.9 \\
  5 & 2.96 & 3.00 & 1.4 & 3.01 & 1.7 \\
  6 & 3.51 & 3.60 & 2.6 & 3.61 & 2.8 \\
  7 & 3.96 & 4.12 & 4.0 & 4.11 & 3.8 \\
  8 & 4.33 & 4.58 & 5.8 & 4.56 & 5.3 \\
  9 & 4.65 & 4.99 & 7.3 & 4.94 & 6.2 \\
  \colrule
  Relative error &   & \multicolumn{ 2}{c}{$ 1 \cdot 10^{-3}$} & \multicolumn{ 2}{c}{$1\cdot10^{-3}$}  \\
  \colrule
  \# terms in PCE &   & \multicolumn{ 2}{c}{2,024} & \multicolumn{ 2}{c}{138} \\
  
  \# FE runs &  53,240  & \multicolumn{ 2}{c}{3,724} & \multicolumn{ 2}{c}{450} \\
  \botrule
\end{tabular}  }
\label{tab:Frame:Reliab}
\end{table} 

From the results in Table~\ref{tab:Frame:Reliab} it is seen that the
sparse polynomial chaos expansion allows one to evaluate accurately
reliability indices up to $\beta=4$ with less than 5\% error. The
parametric analysis w.r.t the threshold is obtained for free since the
PC expansion is computed once and for all. As a whole the analysis has
required 450 calls to the finite element model. Note that it also yields
the statistical moments of the response as well as sensitivity results
at the same cost (see \citet{BlatmanRESS2010} for details).

It should be emphasized that PC expansions provide an approximation
whose accuracy is controlled globally, \ie in a least-square sense. It
does not guarantee a perfect control of the accuracy in the tail of the
distribution of the limit state function though. This means that this
approach should not be used for very small probabilities. Results
reported in the literature show that it is robust up to
$P_f=10^{-4/-5}$.

\subsection{System reliability - meta-IS \citep{DubourgThesis}}
\label{sec:51}
As an illustration consider the following system limit state function
\citep{Waarts2000}:
\begin{equation}
  \label{eq:053}
  g(\Ve{x}) = \min \prt{
    \begin{array}[c]{c}
      3+(x_1- x_2)^2/10 - (x_1 + x_2)/\sqrt 2 \\
      3+(x_1- x_2)^2/10 + (x_1 + x_2) /\sqrt 2 \\
      x_1- x_2 + 7/\sqrt 2 \\
      x_2- x_1 + 7/\sqrt 2
    \end{array}
  }
\end{equation}
where the two components of the input random vector $\Ve{X} $
are independent standard normal variables.

\begin{table}[!ht] 
  \tbl{Meta-importance sampling -- system reliability analysis (after \citet{DubourgThesis})}
  {\tabcolsep12pt
\begin{tabular}{cccc}
  \hline
  & Monte Carlo & Subset simulation & Meta-IS \\\hline 
  Computational cost & 172,000& 284,195 &40 + 200\\
  $\hat{P}_f$ & $2.26 \; 10^{-3} $ &  $2.28 \; 10^{-3} $ &  $2.38\; 10^{-3} $ \\
  C.o.V & $<$ 5\% & $ <$3\% & $<$ 5\% \\
  \hline
\end{tabular}}\label{tab:metaISWaarts}
\end{table} 

The reference results are obtained by crude Monte Carlo simulation so as
to obtain a coefficient of variation less than 5\%. They are reported in
Table~\ref{tab:metaISWaarts} together with the results obtained by
meta-importance sampling. It is shown that only 40~runs of the limit
state function are required in order to build a sufficiently accurate
surrogate. Then only 200 runs of the true model are required using the
sub-optimal instrumental density in order to compute the probability of
failure within 5\% accuracy. The computational cost is thus decreased by
two order of magnitude.

%%%%%%%%%%%%%%%%%%%%%%%%%%%%%%%%%%%%%%%%%%%%%%%%%%%%%%%%%%%%%%%%%%%%%%%%%%%%%%%%%%%%%%%%%%%%%%%%%
\section{Conclusions}
\label{sec:6}
Meta-modelling techniques have become inescapable in modern engineering
since they allow one to address real-world reliability problems at an
affordable computational cost. In this paper, well-established
reliability methods such as FOSM and FORM/SORM are reinterpreted as
basic meta-models. More advanced techniques such as polynomial chaos
expansions and kriging have been reviewed and show promising features in
terms of performance and accuracy.

This paper also emphasizes the need for engineers to catch up with the
latest technologies developed in computer science and applied
mathematics (including statistics) in order to further innovate in their
respective fields.

%\section*{References}

\end{document}